\documentclass{llncs}

\usepackage{booktabs}
\usepackage{braket}
\usepackage{color}
\usepackage{epstopdf}
\usepackage{amsmath}
\usepackage[caption=no]{subfig}
\usepackage{tabularx}
\usepackage{tikz}
\usetikzlibrary{shapes.geometric}

\begin{document}

\title{Quantum circuits for floating-point arithmetic}
\author{Thomas Haener$^{1,3}$, Mathias Soeken$^2$, Martin Roetteler$^3$, and Krysta M. Svore$^3$}
\institute{%
  $^1$ETH Z\"urich, Z\"urich, Switzerland \\
  $^2$EPFL, Lausanne, Switzerland \\
  $^3$Microsoft, Redmond, WA, USA
}
\maketitle

\begin{abstract}
  Quantum algorithms to solve practical problems in quantum chemistry,
  materials science, and matrix inversion often involve a significant
  amount of arithmetic operations which act on a superposition of inputs. These have to be compiled to a set of fault-tolerant low-level operations and throughout this translation process, the compiler aims to come
  close to the Pareto-optimal front between the number of required qubits and the depth of the resulting circuit. In this paper, we provide quantum circuits
  for floating-point addition and multiplication which we find using two vastly different approaches. The first approach is to automatically generate circuits from classical Verilog
  implementations using synthesis tools and the second is to generate and optimize these circuits by hand. We compare our two approaches and provide evidence that floating-point arithmetic is a viable candidate for use in quantum computing, at least for typical scientific applications, where addition operations usually do not dominate the computation.
  All our circuits were
  constructed and tested using the software tools
  LIQ$Ui|\rangle{}$ and RevKit.
\end{abstract}

\section{Introduction}
Quantum computing shows great promise for solving classically intractable computational problems. The wide range of potential applications includes factoring~\cite{shor1994algorithms}, quantum chemistry~\cite{babbush2016exponentially,reiher2016elucidating}, and linear systems of equations~\cite{harrow2009quantum}.
Most of these quantum algorithms invoke subroutines which carry out a classical computation on a superposition of exponentially many input states. Examples include modular exponentiation for factoring~\cite{shor1994algorithms}, evaluating orbital functions for quantum chemistry (e.g., linear combinations of Gaussians)~\cite{babbush2016exponentially}, and reciprocals for solving systems of linear equations~\cite{harrow2009quantum}. While large-scale quantum computers able to run such algorithms are not yet available, it is nevertheless crucial to analyze the resulting circuits in order to acquire resource estimates. These can then guide further development of both quantum algorithms and hardware, allowing for efficient hardware-software co-design.

For instance, recent quantum algorithms to simulate quantum chemistry models implement time evolution under the electronic structure Hamiltonian of electrons interacting with nuclei and with each other. Mathematically, this is described by the Hamiltonian
\[
H = - \sum_i \nabla^2_i/2 -\sum_{i,j} \frac{Z_i}{|R_i - r_j|} + \sum_{j>i} \frac{1}{|r_i - r_j|},
\]
where $\nabla^2_i$ is the Laplace operator of electron $i$, the vectors $R_i$ denote the locations of the nuclei, the scalars $Z_i$ denote the charges of the nuclei, and the $r_i$ are vectors describing the locations of the electrons. When calculating the two-electron integrals of the Hamiltonian $H$ in order to compute the representation of $H$ in second quantization on-the-fly, expressions such as $1/|r_i - r_j|$ must be evaluated. In recent approaches such as \cite{babbush2016exponentially} these expressions would have to be evaluated in superposition, i.e., a quantum circuit is required that computes for instance $x \mapsto 1/x$, where $x$ is a representation of the underlying distances. Also the inverse square root operator $x \mapsto 1/\sqrt{x}$ naturally appears in these computations as typically a computation of Euclidean distances is involved. Several choices seem possible to represent inputs and outputs to these operations, including fixed-point and floating-point representations. Here, we focus on studying the impact of different choices of floating-point representations on the number of quantum bits (qubits) and number of $T$-gates that are required for basic arithmetic operations such as addition and multiplication, from which then more involved operations such as $1/x$ and $1/\sqrt{x}$ can be built, e.g., via Newton's method. 

Compared to a fixed-point representation, floating-point arithmetic offers great savings in number of qubits when the required range of values and/or relative precision is large. Thus, finding good circuits for floating-point arithmetic could be of tremendous use in many applications of quantum computing.

This paper is organized as follows: After a short introduction to floating-point arithmetic and quantum circuits in Sect.~\ref{sec:addmul} and Sect.~\ref{sec:qcirc}, we employ state-of-the-art synthesis tools to transform classical, non IEEE-compliant Verilog implementations to optimized reversible circuits and present the results in Sect.~\ref{sec:synth}. We then introduce our hand-optimized circuits in Sect.~\ref{sec:hand} and mention advantages and disadvantages of using an automatic synthesis approach as opposed to optimizing by hand in Sect.~\ref{sec:advdisadv}. Finally, in Sect.~\ref{sec:viability}, we provide evidence for the viability of using floating-point arithmetic in quantum computing and give a summary and outlook in Sect.~\ref{sec:outlook}.

\section{Floating-point addition and multiplication}\label{sec:addmul}

In a floating-point representation, every number $x$ is approximated using three registers: $1$ sign bit $x_S$, $M$ bits for the (non-negative) mantissa $x_M$ (a number in $[1,2)$), and $E$ bits for the exponent $x_E$. Then,
\[
	x\approx (-1)^{x_S} x_M \cdot 2^{x_E}\;
\]
and because $x_M\in [1,2)$, its highest bit is always 1 and therefore it need not be stored explicitly.
This format allows to represent a much larger range of values with a given number of bits than a fixed-point representation. Yet, basic arithmetic operations require more gates due to the extra steps involved to align and re-normalize intermediate results.
In particular, adding two floating-point numbers $x=(x_S,x_M,x_E)$ and $y=(y_S,y_M,y_E)$ involves the following steps:
\begin{enumerate}
	\item If $x_E < y_E$, swap the two floating-point numbers.
	\item Compute two's complement from sign bits and mantissas (including the implicit leading $1$).
	\item Align the two results according to the difference in exponents $\Delta_E=x_E-y_E$ (only if $|\Delta_E| < M$, else the addition will have no effect).
	\item Add mantissas in two's complement.
	\item Translate from two's complement to sign bit and mantissa.
	\item If adding the two mantissas caused an overflow, right-shift the result by $1$ and increment the exponent.
	\item Determine the position of the first $1$. Left-shift the mantissa by that amount and then update the exponent accordingly.
	\item When copying out the result, check if there was over/underflow in the exponent and output infinity or 0, respectively.
\end{enumerate}
Multiplying two floating-point numbers $x$ and $y$, on the other hand, is much simpler because there is only one renormalization step involved. In summary, it requires the following steps:
\begin{enumerate}
	\item Determine result exponent $r_E=x_E+y_E$.
	\item Multiply mantissas (including the implicit leading $1$) into a $2M$-bit register.
	\item If there was overflow, right-shift by $1$ and increment the result exponent.
	\item If $x_E<0$ and $y_E<0$ but $r_E>0$, output 0.
	\item If $x_E\geq0$ and $y_E\geq0$ but $r_E<0$, output infinity.
	\item Determine the sign bit of the result.
\end{enumerate}
While both floating-point operations are more expensive than their fixed-point analog, the overhead is much more prominent for addition. This fact will have important consequences for our discussion about the practicality of floating-point representations in quantum computing, which we will present in Sect.~\ref{sec:viability}.

\section{Quantum circuits}\label{sec:qcirc}

Programs which run on a quantum computer can be described using quantum circuit diagrams, similar to the one depicted in Fig.~\ref{fig:lut-to-stgb}. Each line represents a qubit and the program is executed from left to right. Because the time evolution of a closed quantum system is described by a unitary operator, every quantum instruction must be reversible (note that this does not include measurement). In particular, executing classical functions on a quantum computer requires mapping all classical non-reversible gates to Toffoli gates (doubly-controlled NOTs) acting on quantum bits. Furthermore, intermediate results need to be stored in temporary work qubits (ancilla qubits) in order to render the computation reversible.

Once the program has been compiled for a certain hardware architecture, the resulting instructions can be executed on the target device. However, physical implementations of qubits are far from perfect and the resulting noise would corrupt the output of any quantum program of nontrivial size. This problem can be remedied by employing quantum error correction which encodes a single logical qubit using many physical ones and thereby reduces the effects of noise on the computation. In order to use quantum error correction, however, all quantum operations need to be mapped to a discrete gate set. One such set of operations is called Clifford+$T$, where the $T$-gate is usually the most expensive quantum operation. There are several proposals to implement a $T$-gate, and all of them feature a large overhead in terms of physical qubits. By, e.g., having many $T$-gate factories available, the runtime of a quantum program can be estimated from the $T$-depth. To estimate the overhead in $T$-gate factories, also the number of $T$-gates which must be executed in parallel is an important measure. In combination with the number of logical qubits, these measures typically allow for a good estimate of the overall cost. We therefore provide these measures for all our circuits. In addition, we provide the circuit size~\cite{steane2003overhead}
\[
	KQ = \text{$T$-depth}\cdot\text{$\#$Qubits}\;,
\]
which can be used to compare different implementations.

\section{Automatic circuit synthesis}\label{sec:synth}
\begin{figure}[t]
  \centering
  \subfloat[\label{fig:lut-to-stga}LUT network]{\begin{tikzpicture}[scale=.25,font=\footnotesize]
  \begin{scope}[every node/.style={inner sep=.5pt}]
    \node (y2) at (243.0bp,306.0bp) [] {$y_2$};
    \node (y1) at (135.0bp,306.0bp) [] {$y_1$};
    \node (x2) at (171.0bp,18.0bp) [] {$x_3$};
    \node (x3) at (99.0bp,18.0bp) [] {$x_2$};
    \node (x1) at (27.0bp,18.0bp) [] {$x_1$};
    \node (x4) at (243.0bp,18.0bp) [] {$x_4$};
    \node (x5) at (315.0bp,18.0bp) [] {$x_5$};
  \end{scope}
  \begin{scope}[every node/.style={draw,ellipse,inner sep=.5pt,minimum width=1.5em}]
    \node (1) at (135.0bp,90.0bp) [] {$1$};
    \node (3) at (135.0bp,162.0bp) [] {$3$};
    \node (2) at (243.0bp,90.0bp) [] {$2$};
    \node (5) at (243.0bp,234.0bp) [] {$5$};
    \node (4) at (243.0bp,162.0bp) [] {$4$};
  \end{scope}
  \draw [->] (x3) ..controls (111.71bp,43.717bp) and (117.15bp,54.286bp)  .. (1);
  \draw [->] (x5) ..controls (290.25bp,43.062bp) and (276.48bp,56.454bp)  .. (2);
  \draw [->] (x1) ..controls (57.672bp,59.328bp) and (94.948bp,108.34bp)  .. (3);
  \draw [->] (x2) ..controls (158.29bp,43.717bp) and (152.85bp,54.286bp)  .. (1);
  \draw [->] (4) ..controls (243.0bp,188.02bp) and (243.0bp,197.29bp)  .. (5);
  \draw [->] (x5) ..controls (309.87bp,66.24bp) and (300.49bp,130.32bp)  .. (279.0bp,180.0bp) .. controls (274.56bp,190.26bp) and (268.06bp,200.63bp)  .. (5);
  \draw [->] (1) ..controls (171.0bp,114.33bp) and (196.62bp,130.94bp)  .. (4);
  \draw [->] (3) ..controls (135.0bp,204.33bp) and (135.0bp,248.79bp)  .. (y1);
  \draw [->] (x4) ..controls (243.0bp,44.017bp) and (243.0bp,53.288bp)  .. (2);
  \draw [->] (1) ..controls (135.0bp,116.02bp) and (135.0bp,125.29bp)  .. (3);
  \draw [->] (5) ..controls (243.0bp,260.02bp) and (243.0bp,269.29bp)  .. (y2);
  \draw [->] (2) ..controls (243.0bp,116.02bp) and (243.0bp,125.29bp)  .. (4);
\end{tikzpicture}

}
  \hfil
  \subfloat[\label{fig:lut-to-stgb}Reversible network]{\begin{tikzpicture}[scale=0.500000,x=1pt,y=1pt,font=\footnotesize]
\filldraw[color=white] (0.000000, -7.500000) rectangle (180.000000, 127.500000);
\draw[color=black] (0.000000,120.000000) -- (180.000000,120.000000);
\draw[color=black] (0.000000,120.000000) node[left] {$x_1$};
\draw[color=black] (0.000000,105.000000) -- (180.000000,105.000000);
\draw[color=black] (0.000000,105.000000) node[left] {$x_2$};
\draw[color=black] (0.000000,90.000000) -- (180.000000,90.000000);
\draw[color=black] (0.000000,90.000000) node[left] {$x_3$};
\draw[color=black] (0.000000,75.000000) -- (180.000000,75.000000);
\draw[color=black] (0.000000,75.000000) node[left] {$x_4$};
\draw[color=black] (0.000000,60.000000) -- (180.000000,60.000000);
\draw[color=black] (0.000000,60.000000) node[left] {$x_5$};
\draw[color=black] (0.000000,45.000000) -- (180.000000,45.000000);
\draw[color=black] (0.000000,45.000000) node[left] {$0$};
\draw[color=black] (0.000000,30.000000) -- (180.000000,30.000000);
\draw[color=black] (0.000000,30.000000) node[left] {$0$};
\draw[color=black] (0.000000,15.000000) -- (180.000000,15.000000);
\draw[color=black] (0.000000,15.000000) node[left] {$0$};
\draw[color=black] (0.000000,0.000000) -- (180.000000,0.000000);
\draw[color=black] (0.000000,0.000000) node[left] {$0$};
\draw (12.000000,105.000000) -- (12.000000,45.000000);
\begin{scope}[rounded corners=2pt]
\begin{scope}
\draw[fill=white] (12.000000, 97.500000) +(-45.000000:8.485281pt and 19.091883pt) -- +(45.000000:8.485281pt and 19.091883pt) -- +(135.000000:8.485281pt and 19.091883pt) -- +(225.000000:8.485281pt and 19.091883pt) -- cycle;
\clip (12.000000, 97.500000) +(-45.000000:8.485281pt and 19.091883pt) -- +(45.000000:8.485281pt and 19.091883pt) -- +(135.000000:8.485281pt and 19.091883pt) -- +(225.000000:8.485281pt and 19.091883pt) -- cycle;
\draw (12.000000, 97.500000) node {{1}};
\end{scope}
\end{scope}
\begin{scope}
\draw[fill=white] (12.000000, 45.000000) circle(3.000000pt);
\clip (12.000000, 45.000000) circle(3.000000pt);
\draw (9.000000, 45.000000) -- (15.000000, 45.000000);
\draw (12.000000, 42.000000) -- (12.000000, 48.000000);
\end{scope}
\draw (24.000000,75.000000) -- (24.000000,30.000000);
\begin{scope}[rounded corners=2pt]
\begin{scope}
\draw[fill=white] (24.000000, 67.500000) +(-45.000000:8.485281pt and 19.091883pt) -- +(45.000000:8.485281pt and 19.091883pt) -- +(135.000000:8.485281pt and 19.091883pt) -- +(225.000000:8.485281pt and 19.091883pt) -- cycle;
\clip (24.000000, 67.500000) +(-45.000000:8.485281pt and 19.091883pt) -- +(45.000000:8.485281pt and 19.091883pt) -- +(135.000000:8.485281pt and 19.091883pt) -- +(225.000000:8.485281pt and 19.091883pt) -- cycle;
\draw (24.000000, 67.500000) node {{2}};
\end{scope}
\end{scope}
\begin{scope}
\draw[fill=white] (24.000000, 30.000000) circle(3.000000pt);
\clip (24.000000, 30.000000) circle(3.000000pt);
\draw (21.000000, 30.000000) -- (27.000000, 30.000000);
\draw (24.000000, 27.000000) -- (24.000000, 33.000000);
\end{scope}
\draw (48.000000,45.000000) -- (48.000000,15.000000);
\begin{scope}[rounded corners=2pt]
\begin{scope}
\draw[fill=white] (48.000000, 37.500000) +(-45.000000:8.485281pt and 19.091883pt) -- +(45.000000:8.485281pt and 19.091883pt) -- +(135.000000:8.485281pt and 19.091883pt) -- +(225.000000:8.485281pt and 19.091883pt) -- cycle;
\clip (48.000000, 37.500000) +(-45.000000:8.485281pt and 19.091883pt) -- +(45.000000:8.485281pt and 19.091883pt) -- +(135.000000:8.485281pt and 19.091883pt) -- +(225.000000:8.485281pt and 19.091883pt) -- cycle;
\draw (48.000000, 37.500000) node {{4}};
\end{scope}
\end{scope}
\begin{scope}
\draw[fill=white] (48.000000, 15.000000) circle(3.000000pt);
\clip (48.000000, 15.000000) circle(3.000000pt);
\draw (45.000000, 15.000000) -- (51.000000, 15.000000);
\draw (48.000000, 12.000000) -- (48.000000, 18.000000);
\end{scope}
\draw (72.000000,30.000000) -- (72.000000,0.000000);
\begin{scope}[rounded corners=2pt]
\begin{scope}
\draw[fill=white] (72.000000, 22.500000) +(-45.000000:8.485281pt and 19.091883pt) -- +(45.000000:8.485281pt and 19.091883pt) -- +(135.000000:8.485281pt and 19.091883pt) -- +(225.000000:8.485281pt and 19.091883pt) -- cycle;
\clip (72.000000, 22.500000) +(-45.000000:8.485281pt and 19.091883pt) -- +(45.000000:8.485281pt and 19.091883pt) -- +(135.000000:8.485281pt and 19.091883pt) -- +(225.000000:8.485281pt and 19.091883pt) -- cycle;
\draw (72.000000, 22.500000) node {{5}};
\end{scope}
\end{scope}
\begin{scope}
\draw[fill=white] (72.000000, 0.000000) circle(3.000000pt);
\clip (72.000000, 0.000000) circle(3.000000pt);
\draw (69.000000, 0.000000) -- (75.000000, 0.000000);
\draw (72.000000, -3.000000) -- (72.000000, 3.000000);
\end{scope}
\draw (96.000000,45.000000) -- (96.000000,15.000000);
\begin{scope}[rounded corners=2pt]
\begin{scope}
\draw[fill=white] (96.000000, 37.500000) +(-45.000000:8.485281pt and 19.091883pt) -- +(45.000000:8.485281pt and 19.091883pt) -- +(135.000000:8.485281pt and 19.091883pt) -- +(225.000000:8.485281pt and 19.091883pt) -- cycle;
\clip (96.000000, 37.500000) +(-45.000000:8.485281pt and 19.091883pt) -- +(45.000000:8.485281pt and 19.091883pt) -- +(135.000000:8.485281pt and 19.091883pt) -- +(225.000000:8.485281pt and 19.091883pt) -- cycle;
\draw (96.000000, 37.500000) node {{4}};
\end{scope}
\end{scope}
\begin{scope}
\draw[fill=white] (96.000000, 15.000000) circle(3.000000pt);
\clip (96.000000, 15.000000) circle(3.000000pt);
\draw (93.000000, 15.000000) -- (99.000000, 15.000000);
\draw (96.000000, 12.000000) -- (96.000000, 18.000000);
\end{scope}
\draw (120.000000,75.000000) -- (120.000000,30.000000);
\begin{scope}[rounded corners=2pt]
\begin{scope}
\draw[fill=white] (120.000000, 67.500000) +(-45.000000:8.485281pt and 19.091883pt) -- +(45.000000:8.485281pt and 19.091883pt) -- +(135.000000:8.485281pt and 19.091883pt) -- +(225.000000:8.485281pt and 19.091883pt) -- cycle;
\clip (120.000000, 67.500000) +(-45.000000:8.485281pt and 19.091883pt) -- +(45.000000:8.485281pt and 19.091883pt) -- +(135.000000:8.485281pt and 19.091883pt) -- +(225.000000:8.485281pt and 19.091883pt) -- cycle;
\draw (120.000000, 67.500000) node {{2}};
\end{scope}
\end{scope}
\begin{scope}
\draw[fill=white] (120.000000, 30.000000) circle(3.000000pt);
\clip (120.000000, 30.000000) circle(3.000000pt);
\draw (117.000000, 30.000000) -- (123.000000, 30.000000);
\draw (120.000000, 27.000000) -- (120.000000, 33.000000);
\end{scope}
\draw (144.000000,120.000000) -- (144.000000,30.000000);
\begin{scope}[rounded corners=2pt]
\begin{scope}
\draw[fill=white] (144.000000, 82.500000) +(-45.000000:8.485281pt and 61.518290pt) -- +(45.000000:8.485281pt and 61.518290pt) -- +(135.000000:8.485281pt and 61.518290pt) -- +(225.000000:8.485281pt and 61.518290pt) -- cycle;
\clip (144.000000, 82.500000) +(-45.000000:8.485281pt and 61.518290pt) -- +(45.000000:8.485281pt and 61.518290pt) -- +(135.000000:8.485281pt and 61.518290pt) -- +(225.000000:8.485281pt and 61.518290pt) -- cycle;
\draw (144.000000, 82.500000) node {{3}};
\end{scope}
\end{scope}
\draw[color=black,dash pattern=on 2pt off 1pt] (138.000000, 105.000000) -- (150.000000, 105.000000);
\draw[color=black,dash pattern=on 2pt off 1pt] (138.000000, 90.000000) -- (150.000000, 90.000000);
\draw[color=black,dash pattern=on 2pt off 1pt] (138.000000, 75.000000) -- (150.000000, 75.000000);
\draw[color=black,dash pattern=on 2pt off 1pt] (138.000000, 60.000000) -- (150.000000, 60.000000);
\begin{scope}
\draw[fill=white] (144.000000, 30.000000) circle(3.000000pt);
\clip (144.000000, 30.000000) circle(3.000000pt);
\draw (141.000000, 30.000000) -- (147.000000, 30.000000);
\draw (144.000000, 27.000000) -- (144.000000, 33.000000);
\end{scope}
\draw (168.000000,105.000000) -- (168.000000,45.000000);
\begin{scope}[rounded corners=2pt]
\begin{scope}
\draw[fill=white] (168.000000, 97.500000) +(-45.000000:8.485281pt and 19.091883pt) -- +(45.000000:8.485281pt and 19.091883pt) -- +(135.000000:8.485281pt and 19.091883pt) -- +(225.000000:8.485281pt and 19.091883pt) -- cycle;
\clip (168.000000, 97.500000) +(-45.000000:8.485281pt and 19.091883pt) -- +(45.000000:8.485281pt and 19.091883pt) -- +(135.000000:8.485281pt and 19.091883pt) -- +(225.000000:8.485281pt and 19.091883pt) -- cycle;
\draw (168.000000, 97.500000) node {{1}};
\end{scope}
\end{scope}
\begin{scope}
\draw[fill=white] (168.000000, 45.000000) circle(3.000000pt);
\clip (168.000000, 45.000000) circle(3.000000pt);
\draw (165.000000, 45.000000) -- (171.000000, 45.000000);
\draw (168.000000, 42.000000) -- (168.000000, 48.000000);
\end{scope}
\draw[color=black] (180.000000,120.000000) node[right] {$x_1$};
\draw[color=black] (180.000000,105.000000) node[right] {$x_2$};
\draw[color=black] (180.000000,90.000000) node[right] {$x_3$};
\draw[color=black] (180.000000,75.000000) node[right] {$x_4$};
\draw[color=black] (180.000000,60.000000) node[right] {$x_5$};
\draw[color=black] (180.000000,45.000000) node[right] {$0$};
\draw[color=black] (180.000000,30.000000) node[right] {$y_1$};
\draw[color=black] (180.000000,15.000000) node[right] {$0$};
\draw[color=black] (180.000000,0.000000) node[right] {$y_2$};
\end{tikzpicture}

}
  \caption{Translation of LUT networks into reversible networks with
    single-target gates}
  \label{fig:lut-to-stg}
\end{figure}
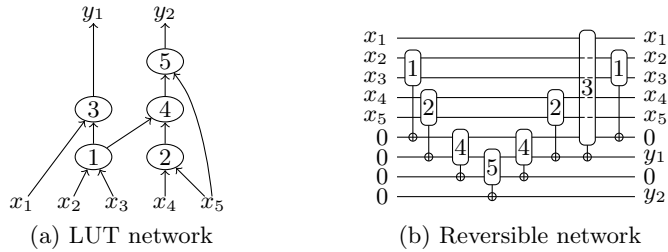
In this section, we present cost estimates for both floating-point
addition and multiplication based on reversible networks that are
obtained from the LUT-based hierarchical synthesis approach
(LHRS,~\cite{SRWM17b}).  LHRS reads as input a classical gate-level
logic network, e.g., provided as Verilog file.  It then uses LUT
mapping techniques (e.g.,~\cite{CD94,CC04,RME+12}) from conventional
logic synthesis algorithms to map the gate-level netlist into a LUT
network composed of $k$-input LUT gates, which can realize any
$k$-input Boolean function.  An example for a LUT network where $k=2$
is illustrated in Fig.~\ref{fig:lut-to-stga}.  Such a network is
translated into a reversible network composed of single-target gates.
Single-target gates are generalized reversible gates in which the
value of a target line is inverted if a given Boolean control function
evaluates to true on the control line values of the
gate. Fig.~\ref{fig:lut-to-stgb} depicts one possible result of such
a translation.  Intermediate values are stored on ancillae, which are
initialized $0$ and need to be restored to their initial value after
computation.  The order in which the LUTs are traversed in this
translation affects the number of required ancillae, because an early
uncomputation of ancilla allows to reuse them for other intermediate
values.  The aim is to find a reversible network with as few ancillae
as possible.  In the reversible network each single-target gates is
mapped to a Clifford+$T$ network.  For this purpose, different
algorithms have been proposed~\cite{ASD15,SRWM17b}.

To obtain circuits using LHRS we first optimized existing IP blocks
for floating-point addition and multiplication for gate count and
mapped them into AND-inverter graphs (AIGs), which are logic networks
that are composed of AND gates and inverters.  We configured the IP
blocks in a way that their functionality is as close to the
functionality of the hand-optimized circuits.  That is, the IP blocks
are not IEEE compliant and rounding is always closest to zero.  The
obtained AIG representation is used as starting point for the initial
$k$-LUT mapping.  As value for $k$, we used the smallest value such
that the number of required qubits does not exceed the number of
qubits obtained from the hand-optimized circuits.  To find that value,
one can run LHRS without mapping the single-target gates into
Clifford+$T$ networks.  This step is typically quite fast, and the
runtime required for it can be neglected.

\begin{table}[t]
  \def\tabcolsep{10pt}
  \footnotesize
  \centering
  \caption{\label{tab:res} Resource counts for the automatically generated circuits.}
  \begin{tabularx}{\linewidth}{lcXrrr}
    \toprule
    Design & Width && Qubits & $T$-count & Runtime \\
    \midrule
    Adder & 16 &&   76 & 112,059,924 &  143.49 \\
    Adder & 16 &&  100 &      40,915 &    2.28 \\
    Adder & 32 &&  239 &      79,415 &   24.22 \\
    Adder & 64 &&  538 &     165,509 &    2.05 \\[4pt]
    Mult  & 16 &&   81 &   3,195,745 & 3276.00 \\
    Mult  & 32 &&  351 &   1,110,133 &  238.82 \\
    Mult  & 64 && 1675 &   7,709,931 & 3318.67 \\
    \bottomrule
  \end{tabularx}
\end{table}

\begin{figure}[t]
  \centering
  \includegraphics[width=0.7\linewidth]{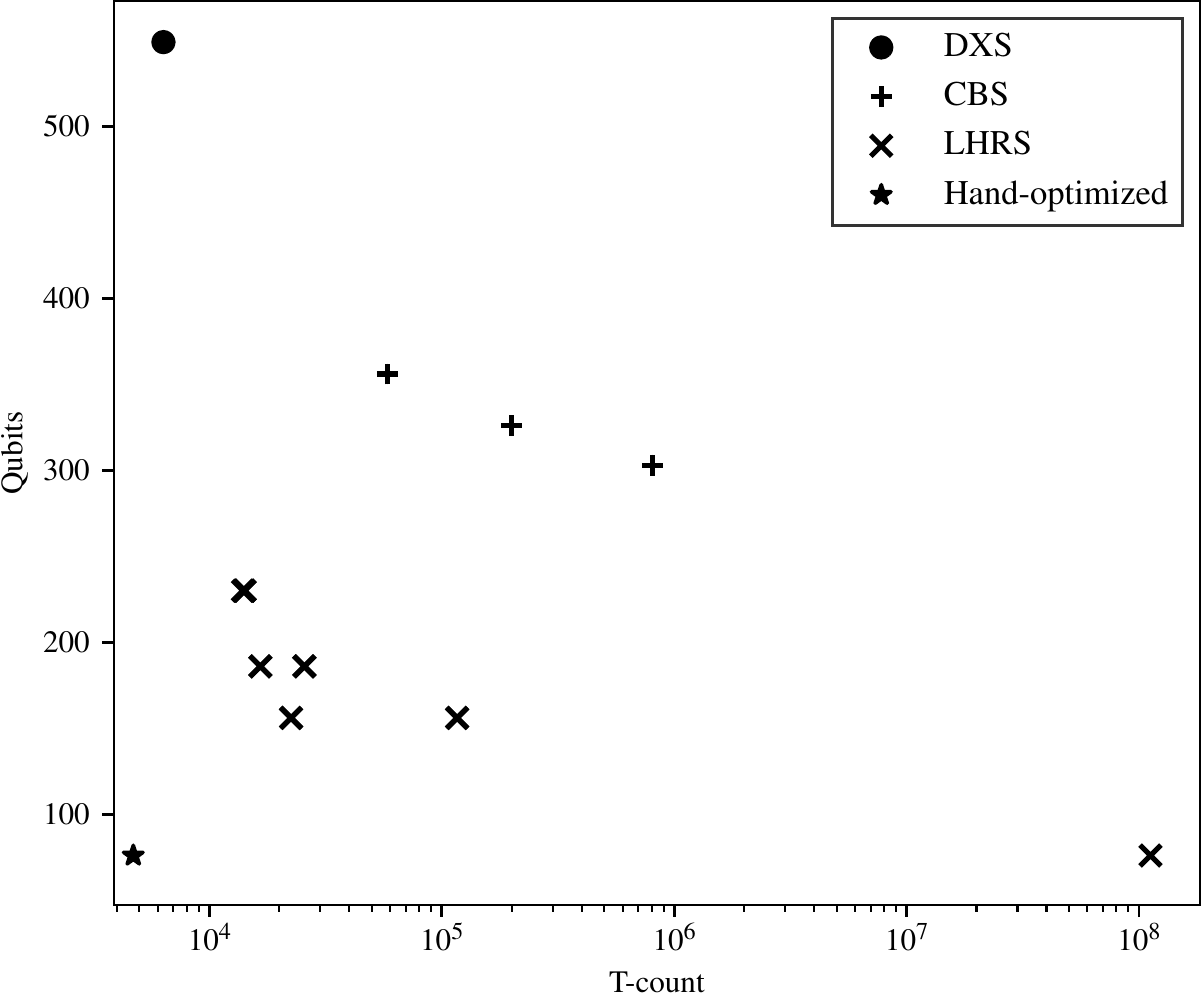}
	\caption{Plot of characteristics of different implementations of $16$-bit floating-point implementations, based on resource counts provided in \cite{SRWM17b}. Each point corresponds to the number of qubits and number of $T$-gates for a particular circuit generated via different circuit synthesis methods, including LHRS. The data point for the hand-crafted circuit is located in the bottom left corner.}
	\label{fig:pareto}
\end{figure}

For each single-target gate, we used all available mappers and
compared the quality of the resulting Clifford+$T$ networks, then
taking the best one. A plot of the parameters of several solutions 
for the case of $16$-bit floating-point adders is shown in Fig.~\ref{fig:pareto}. These circuits, as well as the circuits in Table \ref{tab:res} which improve over some of the results obtained in \cite{SRWM17b} were generated using RevKit, which has implementations of direct XMG-based synthesis (DXS,~\cite{SRWM17}), circuit-based synthesis (CBS,~\cite{SC16}), and LHRS.

\section{Hand-optimized circuits}\label{sec:hand}

In this section, we present hand-optimized circuits for both floating-point addition and multiplication. We detail the individual circuit components and provide resource estimates in order to compare to the synthesis approach discussed in Sect.~\ref{sec:synth}.

\subsection{Basic building blocks}

Our hand-generated floating-point circuits consist of a series of basic building blocks. We use the integer adder from Ref.~\cite{takahashi2009quantum} and construct an integer multiplier from it using the standard shift-and-add approach. To compare two $n$-bit numbers, we perform a subtraction using one extra qubit (i.e., on $n+1$ bits), followed by an addition without this extra qubit, which holds the result of the comparison. If the comparison involves a classically-known constant, we use the CARRY circuit from Ref.~\cite{haner2016factoring}.

The only floating-point-specific blocks are the ones used to determine the location of the first one in a bit-string, and to shift the mantissa by an amount $s$ (specified in an input register). More specifically, the first circuit achieves the mapping
\[
	\Ket x \Ket 0\overset{F}{\mapsto}\Ket x\Ket{\lfloor\log_2(x)\rfloor}\;,
\]
where $x$ is interpreted as a positive integer. The shift circuits $S^\pm$ perform the mapping
\[
	 \Ket s \Ket x \overset{S^\pm}{\mapsto} \Ket s \Ket{2^{\pm s}x} \;.
\]
In this case, $x$ is a $2M$-bit register, where the first/last $M$ bits are guaranteed to be zero, and $s$ is a $\log_2 M$-bit register representing the shift.

\subsection{Implementation}

\begin{figure}[t]
	\centering
	\includegraphics[width=\linewidth]{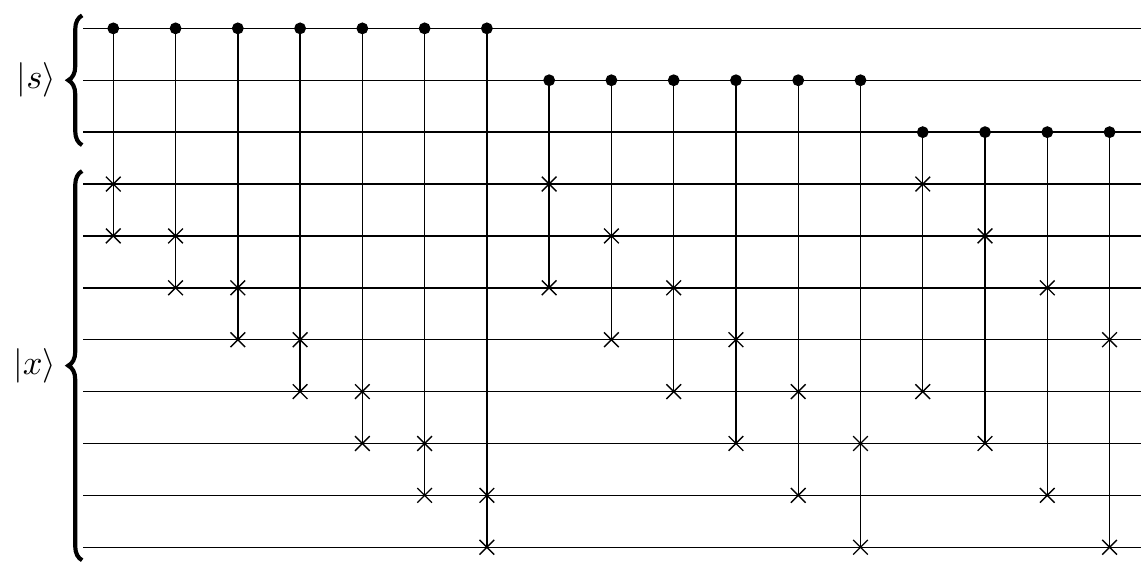}
	\caption{Implementation of a shift circuit for an $8$-bit number $x$. The shift $s$ must be such that the top, i.e., least-significant $s$ bits of $x$ are 0. Variations of this circuit are required to enable shifts in both directions and to ensure that the sign bit is copied when right-shifting a negative number in two's complement.}
	\label{fig:swaps}
\end{figure}

A straight-forward implementation of these shift circuits $S^{\pm}$ would, for every $m\in\{0,...,M-1\}$, copy out the $M$-bit value $x$ shifted by $m$ bits into a new $2M$-bit register, conditional on $s$ being equal to $m$.

To save $M$ qubits, $x$ can first be padded with $M$ bits to the left/right. This allows exchanging the copy-operations above with swaps: For each $m\in\{1,...,M-1\}$, the bits of $x$ can be swapped $m$ bits to the left/right, starting at the left-/right-most bit. Yet, this approach requires $M(M-1)$ Fredkin gates.

A more efficient implementation can be obtained by swapping the bits of $x$ to the left/right by $2^k$, conditional on the $k$-th bit of the shift-register $\Ket s$ and repeating this for every $k\in\{0,...,\log_2 M-1\}$. An example circuit for a $3$-bit shift register and an $8$-bit $x$-register is depicted in Fig.~\ref{fig:swaps}. In general, this circuit requires $\mathcal O(M\log_2M)$ Fredkin gates for a $\log_2 M$-sized shift-register.

\begin{figure}
	\centering
	\includegraphics[width=.9\linewidth]{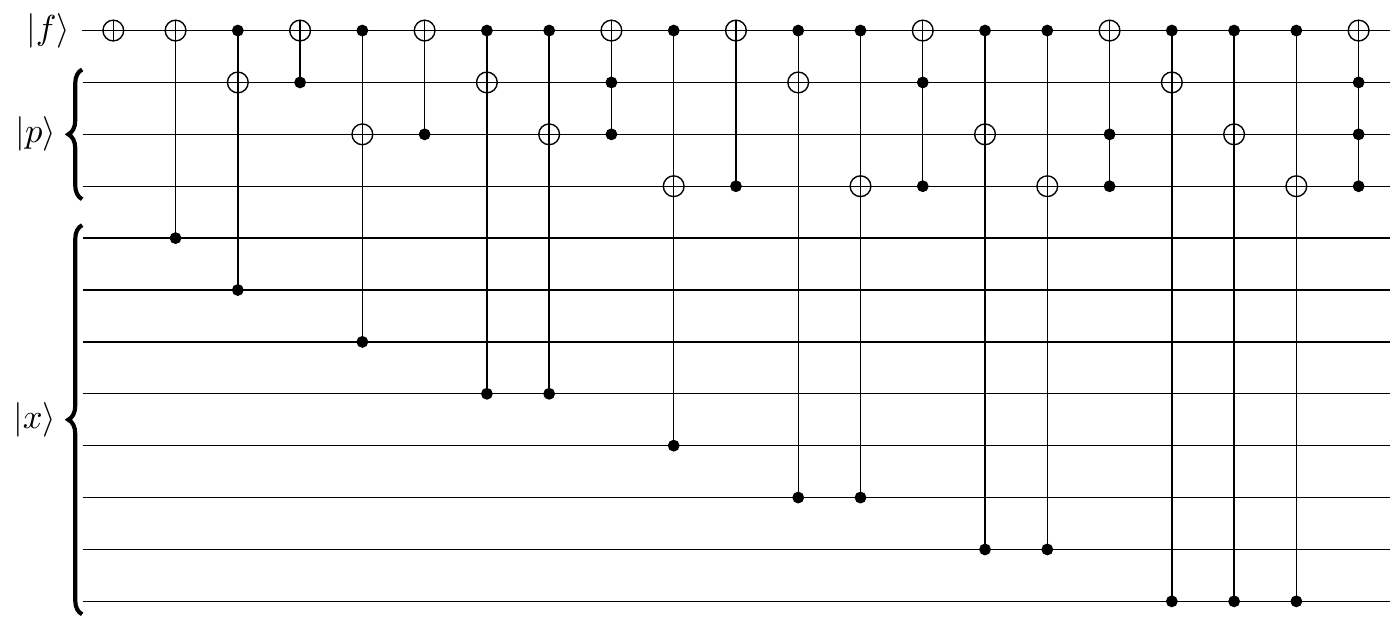}
	\caption{Circuit for finding the first one in the bit-representation of $x$. The flag $f$ (which is initially set to 1 using the first NOT gate) is toggled to 0 as soon as the first 1 has been found. The position of the first one is stored in the $p$-register, consisting of 3 bits in this example.}
	\label{fig:firstone}
\end{figure}

Finding the first one, i.e., implementing the $F$ operation mentioned above, can be achieved using a circuit similar to the one in Fig.~\ref{fig:firstone}, which depicts an example for $8$ bits. The flag $f$ being 1 indicates that the first 1 in the bit-representation of $x$ has not yet been found. For every bit $x_i$ of $x$ (starting with bit-index $i=0$), the position register is initialized to $i$ if the flag is 1 and $x_i=1$ (i.e., it is the first 1). Then, the flag-bit is flipped conditional on the position register being equal to the current $i$ (note that only positive controls need to be placed on the position register).

\begin{figure}
	\centering
	\includegraphics[width=\linewidth]{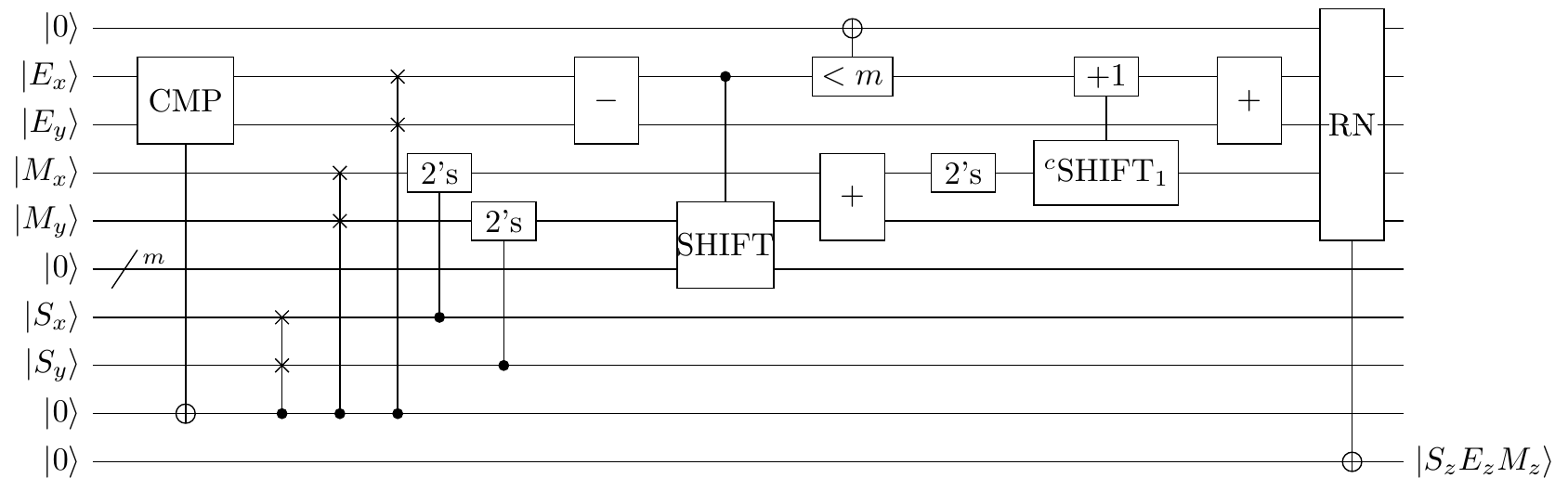}
	\caption{High-level overview of the floating-point addition circuit. First, the inputs are sorted by the exponent (comparison followed by controlled swaps). Then, the second mantissa is shifted by the difference of the exponents before it is added to the first mantissa and converted back from two's complement (taking the pseudo-sign bit as the new sign bit). If there was a final carry in the addition, the result is shifted by 1 bit and the exponent is incremented by 1. The final RN gate renormalizes the intermediate result using the first-ones circuit (see Fig.~\ref{fig:firstone}) followed by shifting the mantissa (see Fig.~\ref{fig:swaps}) by the output of the first-ones circuit and copies out the resulting floating-point representation, taking care of zero and infinity outcomes.}
	\label{fig:add}
\end{figure}

\begin{figure}
	\centering
	\includegraphics[width=.7\linewidth]{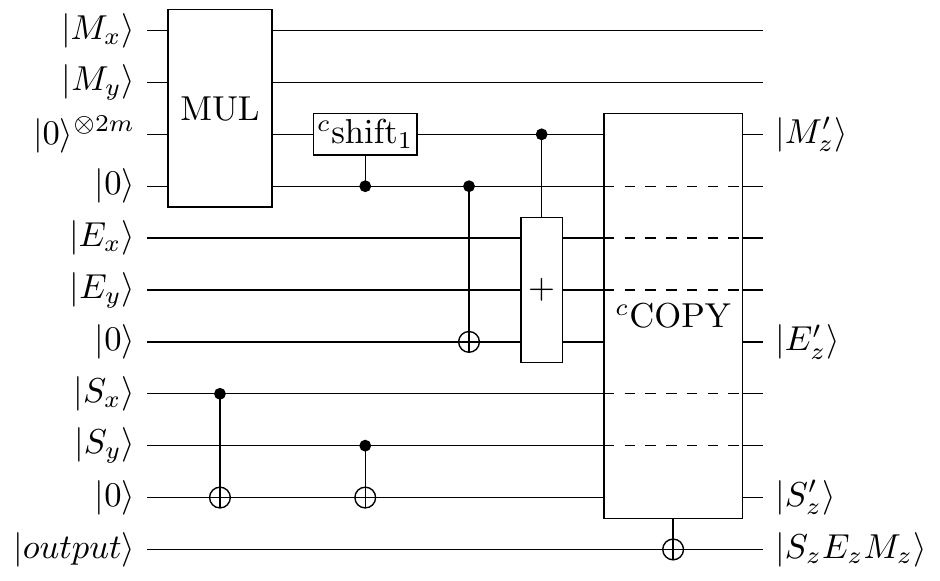}
	\caption{High-level overview of the floating-point multiplication circuit. After multiplying the input mantissas $M_x$ and $M_y$ into a new register of $2m$ qubits ($m$ denotes the number of mantissa bits), it is right-shifted by one if the resulting mantissa $M_x\cdot M_y\geq 2$. The exponent is updated accordingly (using a CNOT), followed by an addition of both input exponents $E_x$ and $E_y$ into $E_z'$ (if $M_x\cdot M_y\neq 0$). The final step denoted by $^c$COPY consists of conditionally copying out of the resulting exponent, mantissa and sign bit to take care of special cases such as over- and underflows in computing the resulting exponent which turn to infinity and zero outcomes, respectively.}
	\label{fig:mult}
\end{figure}

All of the required components were implemented and thoroughly tested using a reversible simulator extension to LIQ$Ui\Ket{}$ \cite{WS14}. The high-level overview circuits for both floating-point addition and multiplication are depicted in Fig.~\ref{fig:add} and Fig.~\ref{fig:mult} and the resource counts which resulted from the implementation in LIQ$Ui\Ket{}$ can be found in Table~\ref{tbl:rescouts}.

\begin{table}[t]
  \def\tabcolsep{10pt}
  \footnotesize
  \centering
  \begin{tabularx}{\linewidth}{lcXrrr}
    \toprule
    Design & Width && Qubits & $T$-count & $T$-depth\\
    \midrule
    Adder & 16 &&  76 &   4,704 & 1,386\\
    Adder & 32 && 140 &  11,144 & 3,138\\
    Adder & 64 && 268 &  26,348 & 7,224\\[4pt]
    Mult  & 16 &&  81 &   6,328 & 2,580\\
    Mult  & 32 && 158 &  26,642 & 11,154\\
    Mult  & 64 && 315 & 122,752 & 52,116\\
    \bottomrule
  \end{tabularx}
  \caption{Resource counts for the hand-optimized circuits. Each Toffoli gate was decomposed using 7 T-gates~\cite{NC00} in $T$-depth 3, providing an upper-bound on the actual T-count~\cite{Jones13}.}
  \label{tbl:rescouts}
\end{table}

\section{Advantages and disadvantages of automatic circuit synthesis}\label{sec:advdisadv}
The results in the previous sections showed that the cost resulting from
the hand-crafted floating-point addition and multiplication circuits are clearly
much lower than the cost resulting from the automatic synthesis tool.  The
main reason for this discrepancy is that the synthesis algorithm is
agnostic to the type of design.  The synthesis approach does not use
the fact that a floating-point operation contains a characteristic
structure, as it is exploited in finding the hand-crafted designs.  It
also highly depends on the logic network that is input to the
synthesis algorithm.  In our case, this has been optimized in order to
reduce the area (in terms of number of gates) in conventional
circuits.  The relation of this objective to the number of qubits and
$T$-count is not fully understood---finding a correlation and deriving
a corresponding cost function from it will significant boost the
effectiveness of the automatic synthesis approach.

Nevertheless, automatic synthesis has clear advantages already in its
current implementation:

\begin{enumerate}
\item One can apply automatic synthesis to various designs and get
  immediate results.  In contrast, to derive a high-quality
  hand-crafted design can require several months.
\item Automatic synthesis can find various different implementations
  of the same design by adjusting the synthesis parameters.  This
  allows for design space exploration.  Depending on the targeted
  quantum platform or the context of the design inside a quantum
  algorithm, one can address different objectives.
\item One may be able to find a design that requires fewer qubits.
  Reducing the number of quantum operations, e.g., $T$ gates, in
  post-synthesis optimization algorithms is much easier than reducing
  the number of qubits.  Automatic synthesis techniques can in
  principle find quantum circuits without any ancilla qubits (except
  to store the result of the outputs).  For example, the 16-bit
  variants of the floating-point adder and multiplier would require
  only 48 qubits.  Although such a circuit is likely to have a very
  large number of quantum operations, the circuit provides a good
  starting point for post-synthesis optimization.
\end{enumerate}

\section{Practicality of floating-point arithmetic for quantum computing}\label{sec:viability}

While the automatic synthesis approach in its current implementation produces very large circuits, floating-point arithmetic for quantum computing is still a viable option, at least when using hand-optimized circuits. Most likely, there are still further improvements possible also in our hand-optimized design: While it features a much lower circuit width than previous adders such as the 32-bit floating-point adder presented in Ref.~\cite{NM14}, the $T$-depth of our design is larger. Specifically, our design requires $1/6$ of the number of qubits of the adder in Ref.~\cite{NM14} and features a size of
\begin{align*}
	KQ &= \text{$T$-depth}\cdot\text{$\#$Qubits}\\
		&\leq 439{,}320\;,
\end{align*}
which is still a $39.3\%$ improvement over the $KQ=723{,}301$ reported in Ref.~\cite{NM14}, despite the much larger $T$-depth.

Furthermore, we argue that exchanging fixed-point arithmetic in a given computation by floating-point arithmetic will result in a circuit of similar cost: While the resource requirements of floating-point addition are much larger than for fixed-point numbers, it is important to note that the cost of floating-point multiplication is very similar to performing it in a fixed-point representation, and given that multiplication in both representations is more expensive than addition, the measure of choice should actually be the cost of multiplication. Furthermore, most applications feature similar numbers of additions and multiplications and often, they can even be combined into a single fused multiply-add instruction which is true, e.g., when evaluating polynomials using the Horner scheme~\cite{knuth1962evaluation}. This means that the overhead of using floating-point arithmetic for applications where multiplications and additions are balanced or where multiplication even dominate is actually much less than what is generally expected. The KQ ratio between a 32-bit floating-point multiplier and a 24-bit fixed-point multiplier (where we require intermediate results to be computed for the full $2M$ bits) is
\begin{align*}
\frac{KQ_{\text{float}}^{\text{mul}}}{KQ_{\text{fixed}}^{\text{mul}}} = \frac{11{,}154 \cdot 158}{10{,}656 \cdot 5 \cdot 24}\approx 1.38\;,
\end{align*}
which clearly shows that the circuit sizes are similar, even for such an unfair comparison: While the chosen bit-sizes guarantee the same absolute precision if no exponent is used, the floating-point multiplier can deal with a much wider range of values at constant relative error. Considering a scientific application with roughly equal numbers of additions and multiplications will cause a deviation from the calculated overhead above by at most another factor of two, since additions require less resources regardless of the chosen representation.

Therefore, we conclude that the cost of using floating-point arithmetic is not only manageable, but that it actually incurs almost no overhead for typical scientific applications. For many quantum algorithms, the extra range and constant relative error offered by a floating-point representation are well worth an increase in circuit size of $2$-$3\times$.

\section{Summary and outlook}\label{sec:outlook}

Given the strict requirements of the IEEE standard, it is expected that IEEE-compliant floating-point arithmetic features large overheads compared to fixed-point arithmetic. Furthermore, even when considering non IEEE-compliant blocks, the number of gates obtained from circuit synthesis is much larger than what would be expected from a fixed-point implementation. Yet, in combination with manual circuit optimization, relaxing the requirements allows for significant savings in both width and size of the circuit, rendering the use of floating-point arithmetic for future quantum devices much more practical. Furthermore, since the cost of multiplying fixed-point numbers is very similar to floating-point multiplication, using floating-point arithmetic in typical scientific applications will incur an overhead in the circuit size $KQ$ of only $2$-$3\times$.

One reason for the large discrepancy between our two approaches---manual optimization and automatic circuit synthesis---is that the objective function used in the optimization process for classical computing is very different from the one used in quantum computing: In classical computing, the most costly resource is time, while bits are essentially free. Circuits resulting from an optimization procedure aiming to minimize the cost function for classical computing are thus highly parallel, but they also require more bits. In quantum computing, on the other hand, both circuit depth and width (i.e., number of bits) are precious resources~\cite{steane2003overhead}. This makes introducing parallelism harder and an optimization procedure would generate vastly different circuits featuring less parallelism and fewer bits.

While the hand-optimized circuits require fewer qubits and $T$-gates, it is very likely that some of the subroutines may still be further optimized using methods from the automatic synthesis approach. Furthermore, the interplay among different components in the hand-written circuit may benefit from such a procedure. We aim to investigate this combination of approaches in future work.

\end{document}